\documentclass[prl,showpacs,preprintnumbers,twocolumn]{revtex4}
\usepackage{amssymb}
\usepackage{color}
\usepackage{graphicx}

\begin{document}

\title[Anomalous Hall effect]{Weak localization correction to the anomalous
Hall effect in polycrystalline Fe films}
\author{P. Mitra, R. Misra, A.F. Hebard}
\email[Corresponding author:~]{afh@phys.ufl.edu}
\author{K.A. Muttalib}
\affiliation{Department of Physics, University of Florida, Gainesville FL 32611}
\author{P. W\"{o}lfle}
\affiliation{ITKM, Universit\"{a}t Karlsruhe, D-76128 Karlsruhe, Germany}
\author{}

\begin{abstract}
\emph{In situ} transport measurements have been made on ultrathin ($<$100 {%
\AA } thick) polycrystalline Fe films as a function of temperature
and magnetic field for a wide range of disorder strengths. For
sheet resistances $R_{xx}$ less than $\sim 3k\Omega$, we find a
logarithmic temperature dependence of the anomalous Hall
conductivity $\sigma _{xy}$ which is shown for the first time to
be due to a universal scale dependent weak localization correction
within the skew scattering model. For higher sheet resistance,
granularity becomes important and the break down of universal
behavior becomes manifest as the prefactors of the $\ln T$
correction term to $\sigma _{xx}$ and $\sigma _{xy}$ decrease at
different rates with increasing disorder.
\end{abstract}

\maketitle



In ferromagnetic metals the anomalous Hall (AH) effect arises,
even in the absence of an applied magnetic field, as a consequence
of the spin-orbit (s-o) interaction of the spin-polarized current
carriers with the non-magnetic periodic lattice and/or impurities.
A full understanding of the AH effect can in principle provide
quantitative estimates of spin-dependent transport coefficients in
magnetic materials. Current understanding is based on several
proposed mechanisms: the skew-scattering model (SSM) \cite{smit},
the side-jump model (SJM) \cite{berger}, and more recently, a
Berry phase model \cite{sundaram}, based on an effect predicted in
the 1950s \cite {karplus} (for a recent review, see
\cite{woelfle}). To calculate the AH conductivity $\sigma
_{xy}^{SSM}$ within the SSM one considers s-o interaction due to
impurity potentials, leading to left/right handedness of the
scattering cross section for electron spin $\uparrow /\downarrow
$. In general, $\sigma _{xy}^{SSM}\varpropto \sigma
_{xx}\varpropto \tau _{tr}$, with $\sigma _{xx}$ the longitudinal
conductivity and $\tau _{tr}$ the transport relaxation time. The
SJM is the consequence of a perpendicular term in the current
density operator that arises from the presence of s-o coupling
induced by impurity potentials.

For the thin ferromagnetic films studied here, disorder is
systematically varied and quantum corrections to $\sigma _{xy}$
become increasingly important. It is known that the normal Hall
conductivity receives quantum corrections from weak localization
(WL), $\delta \sigma _{xy}^{WL}/\sigma _{xy}=2\delta \sigma
_{xx}^{WL}/\sigma _{xx}$ \cite{fukuyama}, but no Coulomb
interaction corrections \cite{altschuler}. Within a weak
scattering model the exchange contribution to the AH conductivity
was found to be zero as well \cite{langenfeld}. In experiments of
Bergmann and Ye (BY)\cite{bergmann} on amorphous Fe films of a few
atomic layers thickness, quantum corrections to the AH
conductivity were found to be negligible. This was interpreted by
the assumption of enhanced spin flip scattering, suppressing the
WL contribution. Theoretically, the WL contribution to the AH
conductivity, although finite in different models
\cite{langenfeld, dugaev}, is cut off by spin-flip scattering
($1/\tau _{s}$), by spin-orbit scattering ($1/\tau _{so}$), by the
magnetic field inside the ferromagnet ($\omega _{H}$), and by the
phase relaxation rate ($1/\tau _{\varphi }$). Whereas $1/\tau
_{s}$ ,\ $1/\tau _{so}$ and $\omega _{H}$ are $T$-independent,
$1/\tau _{\varphi }$ grows linearly with temperature $T$ and leads
to a logarithmic $T$-dependence, provided $\max (1/\tau
_{s},1/\tau _{so},\omega _{H})\ll 1/\tau _{\varphi }\ll 1/\tau
_{tr}$, which defines a temperature interval of observability of
the WL contribution. Unexpectedly, $1/\tau _{\varphi }$ in
ferromagnetic films turns out to be largely due to spin-conserving
inelastic scattering off spin wave excitations
\cite{tatara,plihal}, such that the above inequality is satisfied
in the regime of stronger disorder and WL is experimentally
observed.

We present results on two series of ultrathin films of
polycrystalline iron grown at room temperature by r.f. magnetron
sputtering in the Hall bar geometry through a shadow mask onto
glass (type A samples) and sapphire (type B samples) substrates
under slightly different deposition conditions. The experiments
were performed in a specialized apparatus in which the sample can
be transferred without exposure to air from the high vacuum
deposition chamber to the center of a 7\,T magnet located in a low
temperature cryostat. \textit{Ex situ} topographical scans using
an AFM showed a granular morphology. To parameterize the amount of
disorder in a given film, we use the sheet resistance $R_{0}\equiv
R_{xx}(T=5K)$, which in the results reported here
spans the range from $50\Omega $ (100 {\AA } thick) to $50k\Omega $ ($<$ 20 {%
\AA } thick). Carefully timed postdeposition ion milling of some
of the Fe films gives rise to a decrease in resistance as large as
a factor of two accompanied by a concomitant improvement of
electrical homogeneity and film smoothness \cite{mitra}. This
postdeposition \textit{in situ} treatment does not noticeably
affect the dependence of our transport results on $R_0$, thus
indicating that $R_0$ rather than surface topography is a robust
indicator of disorder. The AH resistivity is measured under
constant current conditions in magnetic fields of $\pm$4T at
selected temperatures. These fields are well above saturation
where the AH signal is maximum. For each sample the longitudinal
and transverse voltages are simultaneously measured so that the
symmetric ($R_{xx}$) and antisymmetric ($R_{xy}$) responses can be
extracted. Contribution from the normal Hall effect is negligible,
as is the magnetoresistance.

We define the \textquotedblleft normalized relative
change\textquotedblright\ , $\Delta
^{N}(Q_{ij})=(1/L_{00}R_{0})(\delta Q_{ij}/Q_{ij})$ with respect
to our reference temperature $T_{0}=5K<T$, where $L_{00}=e^{2}/\pi
h$, $\delta Q_{ij}=Q_{ij}(T)-Q_{ij}(T_{0})$ and $Q_{ij}$ refers to
either resistances $R_{xx},R_{xy}$ or conductances $\sigma
_{xx},\sigma _{xy}$. We find that for all values of
$R_{xx}(T_{0})\equiv R_{0}$ studied, there is a range of
temperatures $T<20K$ where $\Delta ^{N}(Q_{ij})$ has a logarithmic
temperature dependence. Figure 1 shows results for a type A sample
with $R_{0}=2733\Omega $. Following BY's notation \cite{bergmann},
we define for our low temperature data
\begin{equation}
\Delta ^{N}(R_{xx})=-A_{R}\text{ln}\frac{T}{T_{0}};\;\;\;\Delta
^{N}(R_{xy})=-A_{AH}\text{ln}\frac{T}{T_{0}}.
\end{equation}%
Using the fact that $|\delta R_{xx}|\ll R_{0}$, and $R_{xy}(T)\ll
R_{xx}(T)$ for our films, we have the longitudinal conductivity
$\sigma _{xx}\approx 1/R_{xx}$ and the AH conductivity $\sigma
_{xy}\approx R_{xy}/R_{xx}^{2}$, so that

\begin{equation}
\Delta ^{N}\sigma _{xx}=A_{R}\text{ln}\frac{T}{T_{0}};\;\;\;\Delta
^{N}\sigma _{xy}=(2A_{R}-A_{AH})\text{ln}\frac{T}{T_{0}}.
\end{equation}%
Figure 1 also shows that the curves for $\Delta ^{N}(R_{xx})$ and $\Delta
^{N}(R_{xy})$ exactly overlap each other while obeying $\text{ln}T$
dependence up to $T\sim 20K$, i.e. $A_{AH}=A_{R}$.
\begin{figure}[tbp]
\begin{center}
\includegraphics[angle=0, width=0.25\textheight]{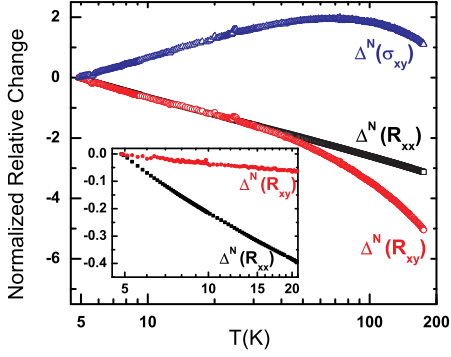}
\end{center}
\caption{$\ln (T)$ dependence of $R_{xx}$ and $R_{xy}$ for $T<20K$
for a type A sample with $R_{0}=2733\Omega $. Inset:
$R_{0}=49k\Omega $.} \label{fig1}
\end{figure}
Fitting the low temperature data to equations (1), we find for this
particular film that $A_{R}=0.897\pm 0.001$, and $A_{AH}=0.908\pm 0.005$.
This `relative resistance' (RR) scaling, namely $A_{AH}\approx A_{R}\approx
1 $, remains valid for type A samples in the range $500\Omega<R_{0}<3k\Omega $
and type B samples in the more restricted range $1.5k\Omega<R_{0}<3k\Omega $, as shown in
Fig.~2.  In contrast, BY observed $A_{R}=1$ and $A_{AH}=2$ within the same
range of $R_{0}$ and $T$. We note from eq (2) that $A_{AH}/A_{R}=2$ implies $%
\Delta^N (\sigma _{xy})=0$. Thus while the \textit{amorphous}
samples of BY
show no logarithmic temperature dependence of AH conductance, our \textit{%
polycrystalline} samples show a $\text{ln}T$ dependence with a
prefactor close to unity. At lower $R_{0}$, $A_{H}$ increases for
both sample types with a more pronounced increase in the type B
samples where $A_{AH}/A_{R}=2$ and $\Delta^N (\sigma _{xy})=0$ at
the lowest resistance $R_{0}=140\Omega$.

At higher resistances, the RR scaling for $T < 20K$ shows
deviations, as seen in the inset of Fig.~1 and in Fig.~2(c). We
will argue later that these deviations can be understood within a
granular model. The deviations from the $\text{ln}T$ behavior at
temperatures $T>20K$ shown in Fig.~1 seem to be non-universal, and
may arise from phonon scattering.
\begin{figure}[bp]
\begin{center}
\includegraphics[angle=0, width=0.3\textheight]{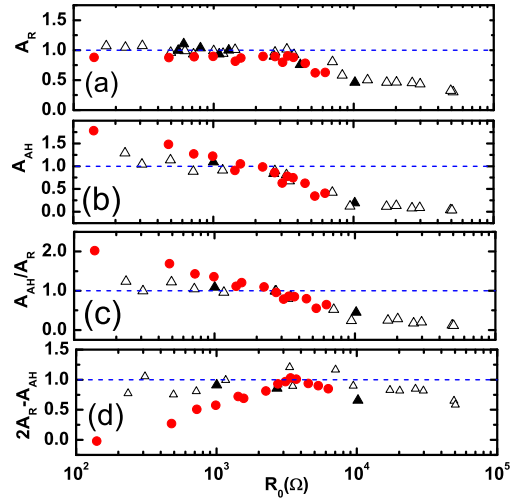}
\end{center}
\caption{Coefficients $A_R$ and $A_{AH}$ as defined in Eq.~1 for
type A (triangles) and type B (circles) samples. Solid triangles
represent type A samples that have undergone surface modification
and conductance changes associated with postdeposition ion milling
as described in text.} \label{fig2}
\end{figure}
This high temperature region could not be studied by BY because their
quenched condensed films irreversibly annealed. We will restrict our
theoretical analysis to the low temperature regime only.

We describe the ferromagnetic film as a quasi-two-dimensional
system of conduction electrons with Fermi energies $\epsilon
_{F\sigma }$ depending on the spin index $\sigma =\uparrow
,\downarrow $, and with spin-orbit coupling $g_{\sigma }$. The
Coulomb interaction will be considered later as a perturbation. We
model the total impurity potential as a sum over identical
single impurity potentials $V(\mathbf{r-R}_{j})$ at random positions $%
\mathbf{R}_{j}$. The Hamiltonian is given by
\begin{eqnarray}
H={\textstyle\sum\limits_{\mathbf{k}\sigma }}(\epsilon
_{\mathbf{k} }&-& \epsilon _{F\sigma })c_{\mathbf{k}\sigma
}^{+}c_{\mathbf{k}\sigma }
+{\textstyle\sum\limits_{\mathbf{k}\sigma ,\mathbf{k}\prime \sigma
\prime , \mathbf{j}}}V(\mathbf{k-k}^{\prime
})e^{i(\mathbf{k-k}^{\prime })\cdot \mathbf{R}_{j}}\cr &\times&
\{\delta _{\sigma \sigma \prime }-ig_{\sigma }\tau _{\sigma \sigma
\prime }^{z}(\widehat{\mathbf{k}}\times \widehat{\mathbf{
k^{\prime }}})\}c_{\mathbf{k\prime }\sigma \prime
}^{+}c_{\mathbf{k}\sigma }
\end{eqnarray}%
where $\tau ^{z}$ is a Pauli matrix, and $\widehat{\mathbf{k}},\widehat{%
\mathbf{k^{\prime }}}$ are unit wave vectors. For simplicity we
assume isotropic band structure and expand the dependence on
scattering angle $\phi $ in angular momentum eigenfunctions
$e^{im\phi }$, keeping only the s-wave component $V$ of the
impurity potential. In terms of the real and imaginary parts of
the $m=1$ eigenvalue of the particle-hole ladder $ \lambda
_{1\sigma }=\lambda _{1}^{\prime }+i\lambda _{1\sigma }^{\prime
\prime }$, we find the spin independent parameters $\tau
_{tr}\equiv \tau /(1-\lambda _{1}^{\prime })$ and $ \overline{\eta
}=\lambda _{1\sigma }^{\prime \prime }/g_{\sigma }\thickapprox
N_{0}V$ where $N_{0}$ is the average DOS at the Fermi level and
$\tau $ is the single particle relaxation time. Keeping only the
main dependencies on spin $\sigma $, that of $\ k_{F\sigma }$, the
longitudinal and the AH conductivities of this model are given in
the weak scattering limit by
\begin{equation}
\sigma _{xy}^{SSM}=\sigma _{xx}^{0}\overline{\eta }Mg_{so}\tau _{tr}/\tau
;\;\;\;\sigma _{xy}^{SJM}=e^{2}Mg_{so}\tau _{tr}/\tau ,
\end{equation}%
where $\sigma _{xx}^{0}=e^{2}(n/m)\tau _{tr}$, $g_{so}=g_{\sigma
}/4\pi n_{\sigma }$ and $M=n_{\uparrow }-n_{\downarrow }$ is the
net spin density. While these equations are consistent with
conventional results $R^{SSM}_{xy}\propto \rho$ and
$R^{SJM}_{xy}\propto \rho^2$, $\rho$ being the resistivity, we
note that the ratio $\sigma^{SJM}_{xy}/\sigma^{SSM}_{xy}$ can in
fact decrease with increasing $R_0$, as we demonstrate later, if
$\bar{\eta}$ increases sufficiently rapidly with disorder,
especially since disorder in our thin films is characterized by
the sheet resistance $R_0$ rather than the resistivity $\rho$.

Turning now to the quantum corrections, we first calculate
\cite{muttalib} that there is no interaction induced $\ln T$
correction to $\sigma _{xy}$. This holds for both exchange and
Hartree terms, for both the skew scattering and the side-jump
models. It generalizes the result reported in \cite{langenfeld}.
To see if WL corrections are important, we need an estimate of the
phase relaxation rate $\tau _{\varphi }^{-1}$ . While the
contribution from e-e-interaction $\tau _{\varphi
}^{-1}=(T/\epsilon _{F}\tau _{tr})\ln (\epsilon _{F}\tau _{tr}/2)$
is small, a much larger contribution is obtained from scattering
off spin waves \cite{tatara,plihal}, $\tau _{\varphi }^{-1}=4\pi
(J^{2}/\epsilon _{F}\Delta _{g})T$, where $J\approx 160K$ is the
exchange energy of the s-electrons and $\Delta _{g}\approx
(1K)(m/m^{\ast })B_{in}$ is the spin-wave gap, where the internal
field $B_{in}$\ \ is in Tesla and $m^{\ast }$ is the effective
mass. For thin films with $150\Omega <R_{0}<3k\Omega $ we find
$\epsilon _{F}\tau _{tr}<10$. Thus the WL condition $\omega
_{H}\tau _{\varphi }<1$\ , with $\omega _{H}=4(\epsilon _{F}\tau
_{tr})(eB_{in}/m^{\ast }c)$, can be satisfied down to $5K$
observing that $B_{in}=B_{ex}$ for a thin slab and taking $
m^{\ast }/m=4$ \cite{hood}. The s-o relaxation rate \ $\tau
_{so}^{-1}\approx g_{\sigma }^{2}/\tau _{tr}\approx 10^{-3}/\tau
_{tr}\approx 0.5-10K$ (using $g_{\sigma }\approx \sigma _{xy}/\sigma _{xx})$%
, which is at the border or below the considered temperature
regime. We have no indication of spin-flip scattering in our
samples. Also, $\tau _{\varphi } $ is seen to be much larger than
$\tau _{tr}$ for all temperatures considered. We are therefore
confident that WL has been seen in our data.

\begin{figure}[tbp]
\begin{center}
\includegraphics[angle=0, width=0.20\textheight]{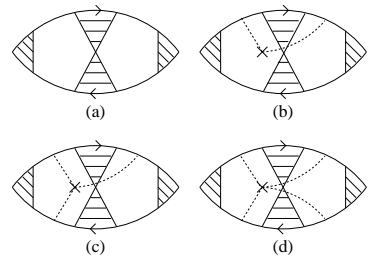}
\end{center}
\caption{Weak localization diagrams. Solid lines are impurity
averaged Green's functions, broken lines are impurity scattering
amplitudes \cite{woelfle,muttalib}), shaded cross is the Cooperon
$ C(Q,i\omega _{l})=(2\pi N_{0}\tau ^{2})^{-1}[|\omega
_{l}|+D^{p}Q^{2}+\tau _{\varphi }^{-1}]^{-1}$ where $D^{p}$ is the
diffusion constant in the particle-particle channel \cite{note1},
and the triangles are current vertex corrections. There are two
diagrams of type (b) and four diagrams of type (c)} \label{WLfig}
\end{figure}

Calculating the diagrams shown in Fig.~3, we find
\begin{eqnarray}
\delta \sigma _{xx}^{WL} = -L_{00}\ln (\tau _{\varphi }/\tau);
\;\;\; \delta \sigma _{xy}^{WL} = \delta \sigma _{xx}^{WL}
\overline{\eta }Mg_{so}\tau _{tr}/\tau .
\end{eqnarray}
The WL contribution from the side-jump model is zero in the
approximation where only the $\sigma -$dependence of \ $k_{F\sigma
}$ is kept and is generally small \cite{dugaev}. Using $\tau
_{\varphi }\sim 1/T$, it follows from Eqs. (4) and (5) that the
normalized WL correction to the Hall conductivity is given as
\begin{equation}
\Delta ^{N}\sigma _{xx}^{WL}=\ln \frac{T}{T_{0}};\;\;\;\Delta
^{N}\sigma _{xy}^{WL}=\frac{\sigma _{xy}^{SSM}\ln
(T/T_{0})}{(\sigma _{xy}^{SSM}+\sigma _{xy}^{SJM})},
\end{equation}
where $\sigma _{xx}^{0}=1/R_{0}$. Comparing with Eq.~(2), this
corresponds to $A_{R}=1$ and $2A_{R}-A_{AH}=[\sigma
_{xy}^{SSM}/(\sigma _{xy}^{SSM}+\sigma _{xy}^{SJM})]$. Assuming
that $\sigma _{xy}^{SJM}/\sigma _{xy}^{SSM}$ is a function of \
$R_{0}$ decreasing from values $\gg 1$ at $ R_{0}\sim 150\Omega $
to values $\ll 1$ at $R_{0}\sim 3k\Omega $, as discussed earlier,
we find good agreement with our experimental data, if $\sigma
_{xy}^{SJM}/\sigma _{xy}^{SSM}$ is small in samples of type A, and
large in samples of type B at weak disorder. The above
interpretation can also account for the BY data on amorphous
films, provided that $\sigma _{xy}^{SJM}/\sigma _{xy}^{SSM}$ is
assumed to be sufficiently large \cite{note3} and the interaction
contribution to $\sigma_{xx}$ is small due to a cancellation of
exchange and Hartree terms  \cite{note2}. One then obtains $\Delta
^{N}\sigma _{xx}=\ln (T/T_{0})$ and $\Delta ^{N}\sigma _{xy}=0$
corresponding to $A_{R}=1$ and $A_{AH}=2$ as seen in the BY
experiment. It is also consistent with data on Fe/Si multilayers
\cite{lin} where the ratio $A_{AH}/A_R$ was found to be
intermediate between BY and RR scaling.

We now turn to the regime $R_{0}>3k\Omega $, where both $A_{R}$
and $A_{AH}$ systematically decrease as $R_{0}$ increases, but at
different rates, as shown in Fig.~2. For example, for a type A
sample with $R_{0}=49k\Omega $, $A_{R}=0.326$ and $ A_{AH}=0.042$,
so that the ratio $A_{AH}/A_{R} \ll 1$. We will now argue that a
granular model for our polycrystalline films explains many of the
qualitative features. As the grains become more weakly coupled the
AH resistivity is dominated by intragranular (rather than
intergranular) skew scattering processes suffered by an electron
when it is multiply reflected off the
grain boundary back into the grain. We may therefore identify $%
R_{xy}=R^g_{xy}$, where $R^g_{xy}$ is the Hall resistivity of a single
grain. Accordingly, in the low resistance regime the longitudinal
resistivity $R_{xx}=R^g_{xx}+R^T_{xx}$ is dominated by $R^g_{xx}$ arising
from the scattering at the grain boundaries while in the high resistance
regime the tunneling process dominates so that $%
R_{xx}\approx R^T_{xx}$. Since $R^g_{xy}$ is independent of
$R^T_{xx}$ in this regime, the quantity $\delta
R_{xy}/R_{xy}=\delta R^g_{xy}/R^g_{xy}$ should be independent of
$R_0$ as is indeed the case as shown in Fig.~4 for type A films
with three different $R_0$.
\begin{figure}[tbp]
\begin{center}
\includegraphics[angle=0, width=0.25\textheight]{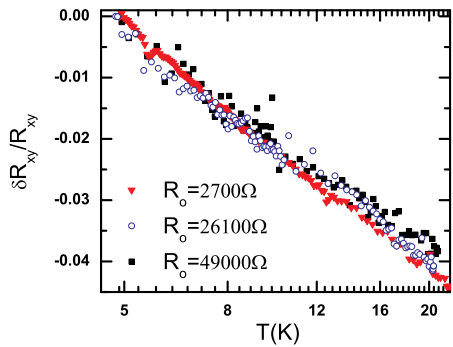}
\end{center}
\caption{$\ln (T)$ dependence of relative changes in the AH
resistance for type A films with three different $R_0$.}
\label{fig4}
\end{figure}

Additional support for the granular model is found by inverting
the resistivity tensor to find $\sigma
_{xy}=R_{xy}^{g}/(R_{xx}^{T})^{2}$. It then follows that $\delta
\sigma _{xy}/\sigma _{xy}=2\delta \sigma _{xx}^{T}/\sigma
_{xx}^{T}+(\delta \sigma _{xy}^{g}/\sigma _{xy}^{g}-2\delta \sigma
_{xx}^{g}/\sigma _{xx}^{g}$). Since our previous calculations
remain valid within a single grain, we can use our calculated
values $A_{R}^{g}=1$
and $A_{AH}^{g}=1$ for the two terms in parentheses. Using $%
\delta\sigma_{xx}^{T}/\sigma_{xx}^{T}=A_{R}R_{0}^{T}L_{00}\text{ln}(T/T_{0})$%
, this yields the relation $\Delta^N (\sigma _{xy}^{WL})=(2A_{R}-R
_{0}^{g}/R _{0}^{T})\text{ln}(T/T_{0})$. Comparing with eq.~(2) we obtain $%
A_{AH}=R_{0}^{g}/R_{0}^{T}$. Since the grain properties are independent of $%
R_{xx}^{T}$, it follows that $A_{AH}\sim 1/R_{0}^{T}$. Figure 2 shows that
this is indeed the case, where $A_{AH}\rightarrow 0$ and the combination $%
2A_{R}-A_{AH}\sim 2A_{R}$ for large $R_{0}$. Although we can not evaluate $%
A_{R}$ separately to explain why $A_R\rightarrow 0.5$ in this regime, we do
expect $A_{AH} $ to be smaller than $A_{R}$, as seen in the inset of Fig.~1
and in Fig.~2, since $A_{R}$ involves only tunnelling resistances while $%
A_{AH}$ involves a ratio of $R_{0}^{g}/R_{0}^{T} \ll 1$. Note,
however, that there is an additional $\ln T$ contribution of the
interaction type due to the scattering of electrons by the
ferromagnetic spin waves: $\delta \sigma _{xx}^{WL}\thicksim
L_{00}(\epsilon _{F}\tau _{tr})^{-2}(J/\epsilon _{F})\ln (\tau
_{\varphi }/\tau )$, which has the opposite sign, and increases
with the disorder strength \cite{muttalib}. This might possibly
lead to a drop of $A_{R}$ beyond $3k\Omega$.

We note that within a granular model, the weak localization
contribution to $\sigma _{xx}$ and the exchange part of the
interaction contribution have been shown to lead to $A_{R}^{WL}=1$
and $A_{R}^{ex}=1$, respectively \cite{belob-lop}. However, the
Hartree contribution to $A_{R}$ has not yet been calculated for
such a model. We expect that it will tend to cancel the exchange
contribution just as in the weakly disordered samples
\cite{note2}. The ``high temperature'' $\ln T$ correction to the
longitudinal resistivity found in \cite{efetov} would appear at
temperatures $T\gtrsim \Gamma $, above the range studied here. We
estimate the inverse escape time $\Gamma =g/\delta $, where $g$ is
the dimensionless conductance and $\delta $ is the average energy
level spacing within a 1~nm granule as $ \Gamma \gtrsim 50K$.

In conclusion, we have investigated the charge transport
properties of ultrathin films of iron grown \textit{in-situ} under
well-controlled conditions, excluding in particular unwanted
oxidation or contamination. Polarizing the magnetic domains by an applied
magnetic field induces a strong anomalous Hall effect signal. We
observe logarithmic temperature dependencies in both the Hall and
the longitudinal conductance over a wide range of temperature and
sheet resistance, which is
a hallmark of quantum corrections in 2D. For sheet resistance \textit{\ }$%
R_{0}<3k\Omega$ in type A samples on glass substrates the
logarithmic corrections obey a heretofore unobserved `RR scaling'
that is found to be nearly independent of $R_{0}$ which is
interpreted in terms of weak localization corrections within the
skew scattering model. In the same regime the type B samples grown
on sapphire substrates show with increasing disorder a continuous
increase of the normalized coefficient of the $\ln T$ term in
$\sigma _{xy}$, which we interpret as a weak localization
correction in the presence of a sizeable but decreasing side-jump
contribution. This interpretation may be extended to the higher
resistance samples, for which the logarithmic corrections are
found to decrease with increasing $R_{0}$, if the granular nature
of these samples is taken into account. Accordingly, it appears
that tunneling resistances between the grains dominate the
longitudinal transport, while the Hall transport is still
controlled by scattering processes within a single grain. Thus the
observed $R_{0}$ dependence of the anomalous Hall effect and its
quantum corrections may be explained by assuming skew scattering
as well as the side-jump mechanism to be operative in our samples.
Quantum corrections of the interaction type turn out to play a
minor role, with the possible exception of an anti localizing
contribution due to scattering off spin waves in the high
resistance regime.

We thank I. Gornyi, D. Maslov and A.D. Mirlin for useful discussions.
This work has been supported by the NSF
under Grant No. 0404962 (AFH), a Max-Planck Research Award (PW, KAM) and by
the DFG-Center for Functional Nanostructures (PW).


\begin{thebibliography}{99}
\bibitem{note1} $D^p$ is equal to the usual diffusion coefficient
$D=\frac{1}{2}v_{F}^{2}\tau _{tr}$ up to terms of order $g_{\sigma
}^{2}$.
\bibitem{note2} The interaction
correction may be small due to strong screening, causing the
exchange and Hartree terms in the interaction correction result
$\delta \sigma _{xx}^{I}=L_{00}(1-\frac{3}{4}\widetilde{F}_{\sigma
})\ln (T\tau )$ to cancel (for bulk Fe one estimates $
\widetilde{F}_{\sigma }\approx 1$ ).
\end{thebibliography}
\end{document}